\documentclass[twocolumn]{pasj02}

\jyear{2025}
\Received{$\langle$reception date$\rangle$}
\Accepted{$\langle$acception date$\rangle$}
\Published{$\langle$publication date$\rangle$}
\usepackage[switch,mathlines]{lineno}

\usepackage{url}
\usepackage{comment}
\usepackage{multirow}
\usepackage{graphicx}	
\usepackage{amsmath}	

\def\kms{km s$^{-1}$\ }

\everymath{\displaystyle}

\begin{document}

\title{Head-tail molecular clouds falling onto the Milky Way disk}

\author{Mikito \textsc{Kohno}\altaffilmark{1,2}\altemailmark,\orcid{0000-0003-1487-5417}\email{kohno.ncsmp@gmail.com}\email{mikito.kohno@gmail.com}
Yasuo \textsc{Fukui}\altaffilmark{2,3},\orcid{0000-0002-8966-9856} 
Takahiro \textsc{Hayakawa}\altaffilmark{2},\orcid{0000-0003-0324-1689} 
Yasuo \textsc{Doi}\altaffilmark{4},\orcid{0000-0001-8746-6548} 
Rin I. \textsc{Yamada}\altaffilmark{3,12},\orcid{0000-0002-1865-4729} 
Fumika \textsc{Demachi}\altaffilmark{2},\orcid{0009-0002-0025-1646} 
Kazuki \textsc{Tokuda}\altaffilmark{5},\orcid{0000-0002-2062-1600} 
Hidetoshi \textsc{Sano}\altaffilmark{3,6},\orcid{0000-0003-2062-5692} 
Shinji \textsc{Fujita}\altaffilmark{7},\orcid{0000-0002-6375-7065} 
Rei \textsc{Enokiya}\altaffilmark{8},\orcid{0000-0003-2735-3239} 
Asao \textsc{Habe}\altaffilmark{9}, 
Kisetsu \textsc{Tsuge}\altaffilmark{3,10,11},\orcid{0000-0002-2794-4840} 
Atsushi \textsc{Nishimura}\altaffilmark{12},\orcid{0000-0003-0732-2937} 
Masato I.N. \textsc{Kobayashi}\altaffilmark{13,14},\orcid{0000-0003-3990-1204} 
Hiroaki \textsc{Yamamoto}\altaffilmark{2},\orcid{0000-0001-5792-3074} and
Kengo \textsc{Tachihara}\altaffilmark{2}\orcid{0000-0002-1411-5410} 
}

\altaffiltext{1}{Curatorial division, Nagoya City Science Museum, 2-17-1 Sakae, Naka-ku, Nagoya, Aichi 460-0008, Japan}
\altaffiltext{2}{Department of Physics, Graduate School of Science, Nagoya University, Furo-cho, Chikusa-ku, Nagoya, Aichi 464-8602, Japan}
\altaffiltext{3}{Faculty of Engineering, Gifu University, 1-1 Yanagido, Gifu, Gifu 501-1193, Japan}
\altaffiltext{4}{Department of Earth Science and Astronomy, Graduate School of Arts and Sciences, The University of Tokyo, 3-8-1 Komaba, Meguro, Tokyo 153-8902, Japan}
\altaffiltext{5}{Faculty of Education, Kagawa University, 1-1 Saiwai-cho, Takamatsu, Kagawa 760-8522, Japan}
\altaffiltext{6}{Center for Space Research and Utilization Promotion (c-SRUP), Gifu University, 1-1 Yanagido, Gifu 501-1193, Japan}
\altaffiltext{7}{Institute of Statistical Mathematics, 10-3 Midori-cho, Tachikawa, Tokyo, Japan}
\altaffiltext{8}{National Astronomical Observatory of Japan (NAOJ), National Institutes of Natural Sciences (NINS), 2-21-1 Osawa, Mitaka, Tokyo 181-8588, Japan}
\altaffiltext{9}{Department of Physics, Faculty of Science, Hokkaido University, N10 W8, Kitaku, Sapporo, Hokkaido 060-0810, Japan}
\altaffiltext{10}{Institute for Advanced Study, Gifu University, 1-1 Yanagido, Gifu, Gifu 501-1193, Japan}
\altaffiltext{11}{Institute for Advanced Research, Nagoya University, Furo-cho, Chikusa-ku, Nagoya, Aichi 464-8601, Japan}
\altaffiltext{12}{Nobeyama Radio Observatory, National Astronomical Observatory of Japan (NAOJ), National Institutes of Natural Sciences (NINS), 462-2, Nobeyama, Minamimaki, Minamisaku, Nagano 384-1305, Japan}
\altaffiltext{13}{I. Physikalisches Institut, Universit\"{a} t zu K\"{o}ln, Z\"{u}lpicher Stra\ss e 77, 50937 K\"{o}ln, Germany}
\altaffiltext{14}{National Institute for Fusion Science (NIFS), 322-6 Oroshi-cho, Toki city, Gifu, 509-5292 Japan}


\maketitle

\begin{abstract}
{We report discovery of two CO clouds which are likely falling down to the Galactic plane at more than $35$ \kms. The clouds show head-tail distributions elongated perpendicular to the Galactic plane at $l=\timeform{331.6D}$ and {$b=\timeform{0D}$} as revealed by an analysis of the Mopra CO $J=$1--0 survey data. We derived the distance of the clouds to be $2.46 \pm 0.18$ kpc based on the Gaia Data Release 3. The CO clouds have molecular masses of $4.8\times 10^3\ M_{\odot}$ and $3.5\times 10^3\ M_{\odot}$, respectively, and show kinetic temperature of 30--50 K as derived from the line intensities of the $^{13}$CO $J$~=~2--1, $^{12}$CO $J$~=~1--0, and $^{13}$CO $J$~=~1--0 emission. The temperature in the heads of the clouds is significantly higher than 10 K of the typical molecular clouds, although no radiative heat source is found inside or close to the clouds. Based on the results, we interpret that the present clouds are falling onto the Milky Way disk and are significantly heated up by the strong shock interaction with the disk HI gas. We suggest that the clouds represent part of the HI intermediate velocity clouds falling to the Galactic plane which were converted into molecular clouds by shock compression. This is the first case of falling CO clouds having direct observed signatures of the falling motion including clear directivity and shock heating. Possible implications of the CO clouds in the evolution of the Galactic interstellar medium are discussed. }
\end{abstract}

\section{Introduction}
{Vertical motions of the interstellar clouds relative to the Galactic plane can link the halo with the Galactic disk, whereas details of the vertical motions are not well understood because of difficulty in detecting motions perpendicular to the line of sight. A population of such clouds falling to the plane are the HI high velocity clouds (HVCs) having blue-shifted radial velocity from 40 \kms to 300 \kms. In particular, the low velocity HVCs in a velocity range of 40-100 \kms are named the intermediate velocity clouds (IVCs), and one of them IVC 86-38 in the Pegasus--Pisces Arch (PP Arch) shows compelling evidence for falling clouds such as the head-tail distribution with a low dust-to-gas ratio \citep{2021PASJ...73S.117F}, which is a unique case of the cloud-halo interaction as supported by numerical simulations of \citet{2022ApJ...925..190S} (see also \cite{2009ApJ...699.1775K}). \citet{2022ApJ...925..190S} predicted that IVC 86-38 having HI mass of $\sim 10^4 M_{\odot}$ will be decelerated by the halo, and {merge with} the Galactic disk in $\sim 10$ Myr, suggesting that most of the IVCs may be accreting {onto} the Galactic disk. 
{While detections of CO lines in IVCs were reported previously (e.g., \cite{1990ApJ...355L..51D,1999A&A...344..955W,2010ApJ...722.1685M,2016A&A...592A.142R}), direct observational signatures of falling motion such as the moving direction and velocity were not revealed.}
We expect formation of H$_2$ and CO in the IVCs due to the high pressure near the plane. {Studying such falling CO clouds on the Milky Way disk will be crucial in revealing the cloud properties and will shed light on the evolution of the IVCs and star formation therein. }

{In the present paper, we report the results obtained by chance in the course of an observational investigation of the giant molecular cloud toward RCW 106 \citep{2025AJ....169..181K}. The RCW 106 (G333) GMC is located at $l=\timeform{330D}$ to $l=\timeform{335D}$ and exhibits active star formation. 
{Toward the region we discovered small CO clouds of $\sim 10^3\ M_{\odot}$ at velocity shifted by $\sim 15$ \kms from the GMC, and the purpose of the paper is to present the detailed observational properties of the clouds.}
The paper is organized as follows; Section 2 describes the datasets employed, Section 3 presents the results of the cloud distribution and their physical properties such as mass and temperature along with distance determination, and Section 4 discussion on the physical implications of the clouds along with the future direction of the research topic. Section 5 concludes the paper.}

\section{Data}

\begin{table*}[h]
\caption{{Properties of CO line data}}
\begin{center}
{
\begin{tabular}{cccccccccc}
\hline 
Telescope & Line  & HPBW & Grid &  Velocity & r.m.s noise$^*$ & References \\
& & &  &Resolution & level &\\
\hline
Mopra &$^{12}$CO~$J$~=~1--0 & $\sim \timeform{33"}$  & \timeform{30"} &0.1 km s$^{-1}$& $\sim 0.8 $ K & [1,2,3,4] \\
 &$^{13}$CO~$J$~=~1--0 &  $\sim \timeform{33"}$ & \timeform{30"} & 0.1 km s$^{-1}$ & $\sim 0.3$ K  & [1,2,3,4] \\
APEX &$^{13}$CO~$J$~=~2--1 &  $\sim \timeform{30"}$ &\timeform{9.5"} & 0.25 km s$^{-1}$ & $\sim 0.8$ K  & [5,6,7] \\
\hline
\end{tabular}}
\label{obs_param}
\end{center}
{\raggedright {Note. $^*$The r.m.s noise level is taken from the final cube data using this paper. 
\\References [1] \citet{2013PASA...30...44B}, [2] \citet{2015PASA...32...20B}, [3] \citet{2018PASA...35...29B}, [4] \citet{2023PASA...40...47C}, [5]\citet{2017A&A...601A.124S}, [6] \citet{2021MNRAS.500.3064S}, [7] \citet{2021MNRAS.500.3027D}}\par}
\end{table*}

\subsection{Mopra Southern Galactic plane CO $J=$1--0 survey}
We analyzed the {CO $J$~=~1--0} data from the Mopra Southern Galactic Plane CO Survey \citep{2013PASA...30...44B,2015PASA...32...20B,2018PASA...35...29B,2023PASA...40...47C}.
Mopra is a 22-m radio telescope operated by the Australia Telescope National Facility (ATNF) in Australia. It observed $^{12}$CO, $^{13}$CO, C$^{18}$O, and C$^{17}$O in the southern galactic plane during the winter seasons of 2011--2018. 
The observations were carried out in a rectangular unit of $60' \times 6'$ using the Fast-On-The-Fly (FOTF) mode.
The front-end utilized a Monolithic Microwave Integrated Circuit (MMIC) receiver operating in the 3 mm band, and the back-end employed the UNSW Mopra Spectrometer (MOPS). 
A part of the survey observations {from June 2018 to November 2018} were performed remotely from Nagoya in Japan through an internet connection by collaborators in the radio astronomy group at Nagoya University. 
The final spatial resolution at 115 GHz with the Mopra telescope is 33\arcsec.
The survey data {were} converted to main-beam temperature using the formula $T_{\rm MB} = T_{\rm A}^*/\eta$, with extended beam efficiency $\eta=0.55$ \citep{2005PASA...22...62L}.
The final spatial resolution of the cube data {archived as Data Release 4} is 36\arcsec, and the velocity resolution is 0.1 \kms.
The cube data of Data release 4 {of} $^{12}$CO and $^{13}$CO $J=$1-0 used in the analysis were downloaded from the web page\footnote{\url{https://mopracosurvey.wordpress.com}}.
The grid size of the data finally used was $(l,b,v)=$(30\arcsec, 30\arcsec, 1 \kms), and the rms noise level was $T_{\rm mb}$ scale of $\sim 0.8$ K for $^{12}$CO $J=$1-0 and $\sim 0.3$ K for $^{13}$CO $J=$1-0, respectively.

\subsection{SEDIGISM: APEX Galactic plane $^{13}$CO $J=$2--1 survey}
We also utilized archived data from the Atacama Pathfinder Experiment telescope (APEX), a 12m submillimeter telescope at an altitude of 5100 m in Chajnantor, Chile. The data were obtained by the SEDIGISM (Structure, Excitation and Dynamics of the Inner Galactic Interstellar Medium: \cite{2017A&A...601A.124S}, \yearcite{2021MNRAS.500.3064S}; \cite{2021MNRAS.500.3027D}) project. The receiver and spectrometer used in the observations were the Swedish Heterodyne Facility Instrument (SHFI) and the wide-band Fast Fourier Transform Spectrometer (XFFTS), respectively. The HPBW is 30\arcsec and the velocity resolution is 0.25 \kms.
The data is calibrated to the $T_{\rm mb}$ scale, the grid size is $(l,b,v)=$(9.5\arcsec, 9.5\arcsec, 0.25\ \kms), and the r.m.s noise level is $\sim 0.8$ K at $T_{\rm mb}$ scale.
{We summarized the properties of CO line data using in this paper in Table \ref{obs_param}.}

\subsection{The Herschel far-infrared data}
We used the archival infrared image data obtained by Herschel\footnote{Herschel Space Observatory is an ESA space observatory with science instruments provided by European-led Principal Investigator consortia and with important participation from NASA \citep{2010A&A...518L...1P}} infrared Galactic Plane Survey (Hi-GAL: \cite{2010PASP..122..314M})
The $160\ \mu$m image data is obtained by the Photodetector Array Camera and Spectrometer (PACS: \cite{2010A&A...518L...1P}). The spatial resolution of $160\ \mu$m image is $12\arcsec$.

\subsection{Gaia Data Release 3}
To estimate the distances to the head--tail molecular clouds, we used stellar data from Gaia Data Release 3 \citep{2023A&A...674A...1G}. We adopted “geometric distances" and G-band extinctions ($A_{\rm G}$) from the catalog of \citet{2021AJ....161..147B}. Uncertainties in $A_{\rm G}$ and geometric distance were taken as the 16th and 84th percentiles, representing the 68\% confidence interval \citep{2021AJ....161..147B,2023A&A...674A...1G}.


\section{Results}
\begin{figure*}[h]
 \includegraphics[width=18cm]{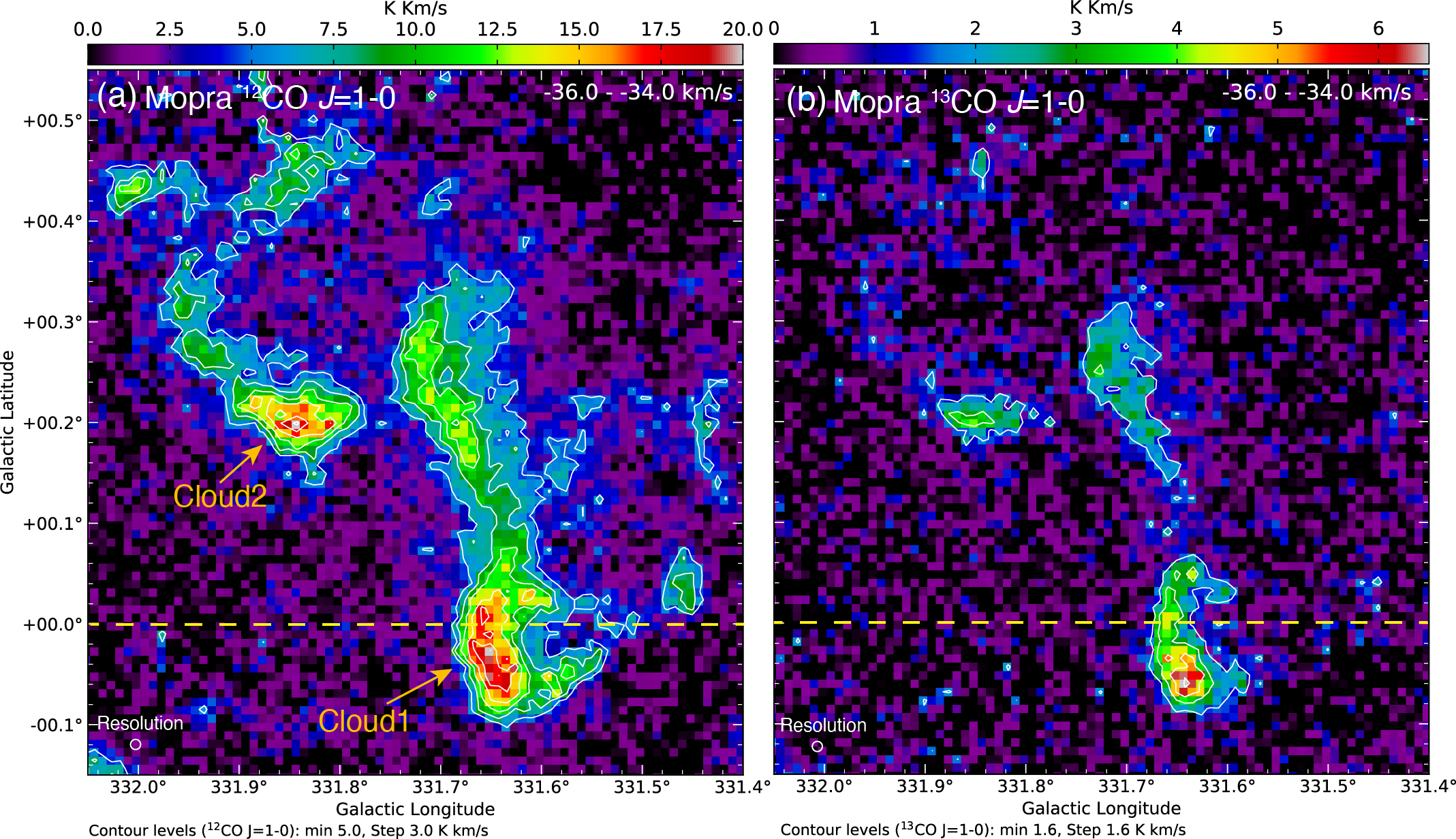}
\caption{(a) The Mopra $^{12}$CO $J$~=~1--0 integrated intensity map of the head-tailed molecular clouds. The lowest contour levels and intervals are 5.0 K \kms and 3.0 K \kms. (b) Same as (a), but for $^{13}$CO $J$~=~1--0. The lowest contour levels and intervals are 1.6 K \kms and 1.6 K \kms. The integrated velocity ranges are from $-36$ \kms to $-34$ \kms. The yellow dotted lines show the Galactic plane ($b=\timeform{0D}$). Alt text: Mopra 12-CO and 13-CO integrated intensity maps.}
\label{mopra}
\end{figure*}

\begin{figure*}[h]
\begin{center} 
 \includegraphics[width=18cm]{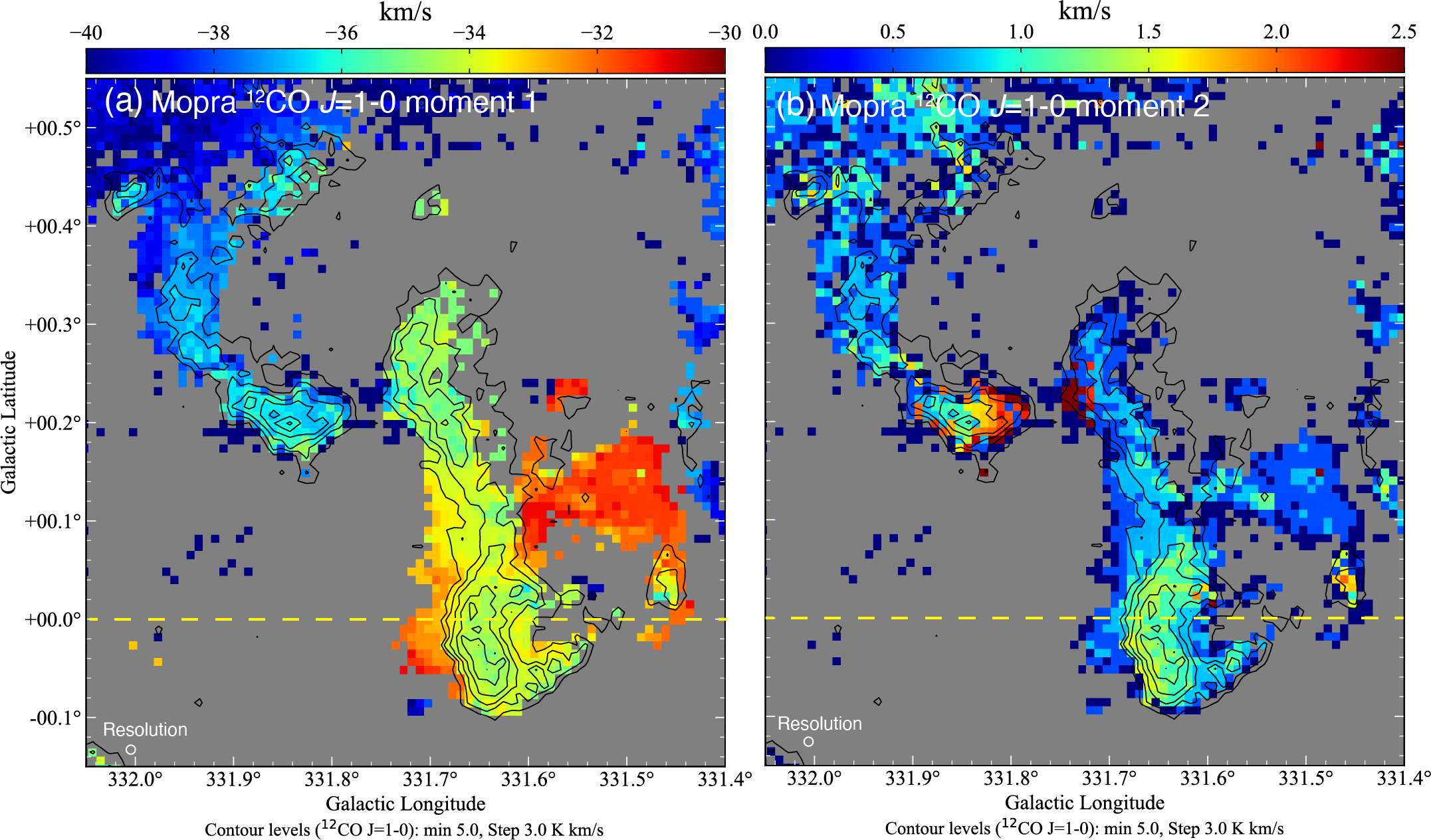}
\end{center}
\caption{The Mopra $^{12}$CO $J$~=~1--0 velocity-field (first-moment) map. (b) The $^{12}$CO $J$~=~1--0 velocity dispersion (second-moment) map. The adopted velocity range extends from $-40$ \kms to $-30$ \kms. The yellow dotted lines show the Galactic plane ($b=\timeform{0D}$). The contour levels and intervals are the same as in Figure \ref{mopra}(a). The data points are plotted above $4.5$ K \kms of the $^{12}$CO $J$~=~1--0 integrated intensity. Alt text: Mopra 12-CO first and second moment maps.}
\label{mom}
\end{figure*}

\subsection{Spatial and velocity distributions of molecular clouds}
{In the course of the investigation of the GMC toward RCW 106 \citep{2025AJ....169..181K}}, we discovered two head-tail molecular clouds elongated perpendicular to the Galactic plane at $l = \timeform{331.6D}$. {The radial velocity of two clouds are separated from the GMC by $\sim 15$ \kms.}
Figures \ref{mopra}(a) and 1(b) show the integrated intensity maps of the head-tail molecular clouds of $^{12}$CO and $^{13}$CO $J$~=~1--0 obtained by Mopra. 
Hereafter, we call these head-tail molecular clouds "Cloud 1" and "Cloud 2", respectively.
Cloud 1 has a peak at the Galactic plane $(l,b)=(\timeform{331.65D},\timeform{-0.03D})$ {(the cloud head)} and extends vertically over $\sim \timeform{0.4D}$ {(the cloud tail). In the distribution of $^{13}$CO, we can find that the peak position of Cloud 1 is $b\sim \timeform{-0.05D}$, which is shifted to the negative $b$ side of the Galactic plane. }
Cloud 2 has a peak at $(l,b)=(\timeform{331.85D},\timeform{+0.20D})$ {(the cloud head)} and has a C-shaped distribution extended toward the positive $b$ side {(the cloud tail)}. 

{We estimated the H$_2$ column density} [$N({\rm H_2})$] from the integrated intensity [$I_{\rm ^{12}CO}$] {of} $^{12}$CO~$J$~=~1--0. $N({\rm H_2})$ is given by {equation (1),}
\begin{equation}
N({\rm H_2}) = X_{\rm CO}\ I_{\rm ^{12}CO}\ {\rm [cm^{-2}]},
\label{XCO}
\end{equation}
where $X_{\rm CO}$ is the CO-to-H$_2$ conversion factor.
In this paper, we used $X_{\rm CO} = 2.0 \times 10^{20}\ [{\rm\ cm^{-2} (K\ km\ s^{-1})^{-1}}]$ of the typical value in the Galactic disk \citep{2013ARA&A..51..207B,2020MNRAS.497.1851S,2024MNRAS.527.9290K}. The peak column densities in the head of Cloud 1 and Cloud 2 are derived to be $N({\rm H_2})\sim 4.0 \times 10^{21}$ cm$^{-2}$ and $N({\rm H_2})\sim 4.2 \times 10^{21}$ cm$^{-2}$, respectively.
{In order to} investigate the velocity distribution of {Cloud 1 and Cloud 2}, we made the intensity-weighted velocity (the first moment: $V_{\rm c}$) and velocity dispersion (the second moment: $\sigma_v$) maps, {by} calculating the equations {(2) and (3),}
\begin{eqnarray}
V_{\rm c} &=& {\int T_{\rm B}(v) \cdot v\ dv \over \int T_{\rm B}(v)\ dv}\ [{\rm km\ s^{-1}}],
\label{eq:vc}\\
\sigma_v &=& \sqrt{{\int T_{\rm B}(v) \cdot (v-V_c)^2 \ dv \over \int T_{\rm B}(v)\ dv}}\ [{\rm km\ s^{-1}}],
\label{eq:dv}
\end{eqnarray}
where $T_{\rm B}$ is the brightness temperature and $v$ is the radial velocity of {the $^{12}$CO $J=$1-0 emission}. If a spectrum has a single Gaussian shape, the relation between $\sigma_v$ and full-width half maximum (FWHM) line-width $\Delta V$ is $\Delta V=\sqrt{8 \ln 2}\ \sigma_v$.
Figures \ref{mom}(a) and \ref{mom}(b) present the first and the second moment maps, respectively.
These moment maps were created for {a} velocity range from $-40$ \kms to $-30$ \kms. 
In Figure \ref{mom} (a), Cloud 1 is distributed with a uniform radial velocity of $\sim -35$ \kms.
Cloud 2 has a velocity range from $-39$ \kms to $-36$ \kms.
We can find that the velocity dispersion of head in Cloud 1 has 1.4--1.5 \kms, which is larger than the tail having $\sim 0.8$ \kms as shown in Figure \ref{mom}(b).
In particular, Cloud 2 has a large velocity dispersion of 2.0--2.5 \kms at the intensity peak {in the head part and a velocity dispersion of 0.3--0.7 \kms in the tail part.}

\subsection{The $^{13}$CO $J$~=~2--1/1--0 intensity ratio}
\begin{figure*}[h]
\begin{center} 
\includegraphics[width=13cm]{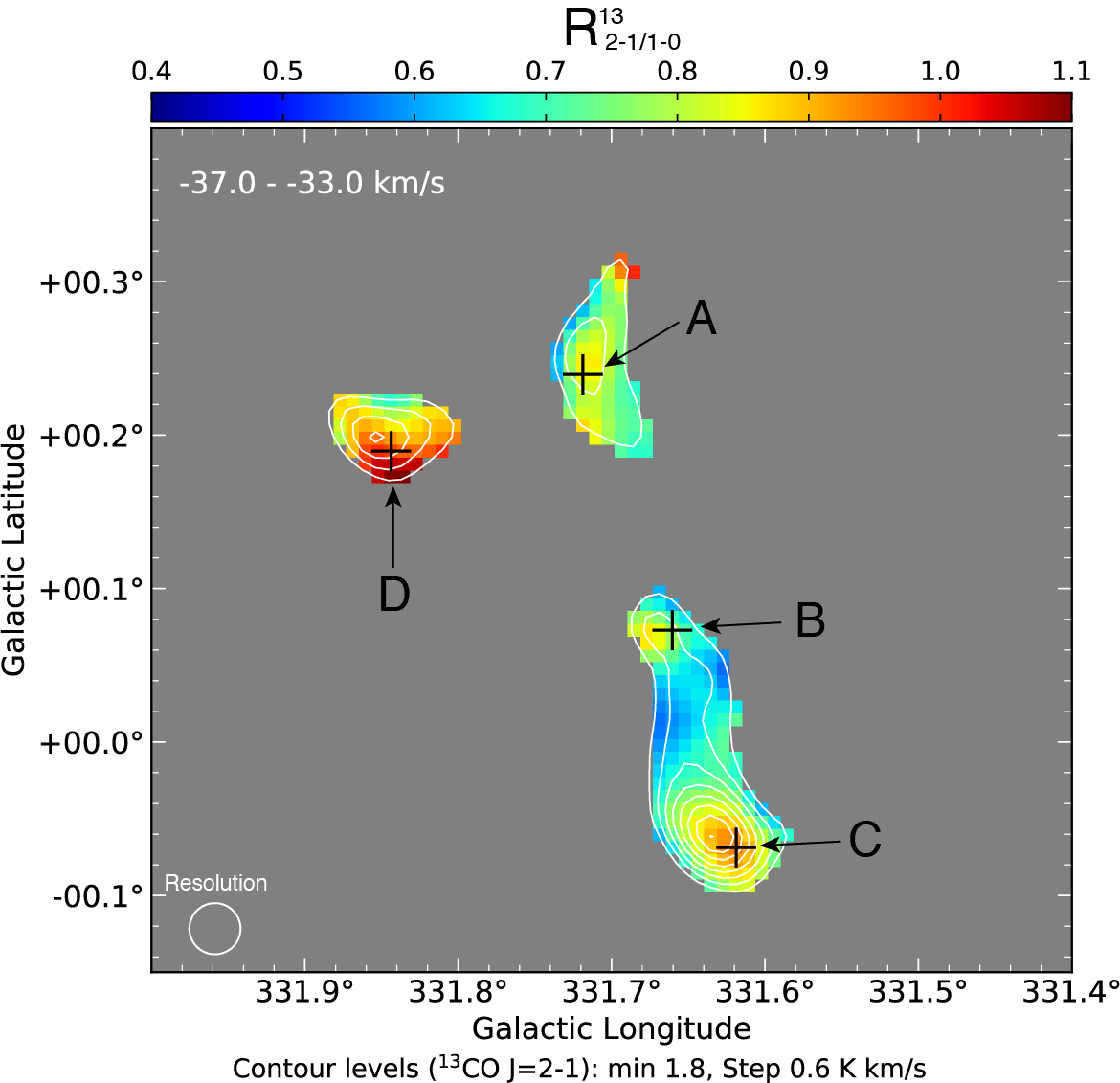}
\end{center}
\caption{The $^{13}$CO $J$~=~2--1/1--0 ratio map above the integrated intensity of 1.8 K \kms. The integrated velocity range is from $-37$ \kms to $-32$ \kms. The data are smoothed to the spatial resolution of \timeform{120"}. The lowest contour level and interval are 1.8 K \kms and 0.6 K \kms of $^{13}$CO $J=$2--1, respectively. A, B, C, and D indicate the position of the LVG calculation  as shown in Figure \ref{lvg} and Table \ref{lvgtable}. Alt text: The 13-CO intensity ratio map.}
\label{ratio}
\end{figure*}
We analyzed the intensity ratio between $^{13}$CO $J$~=~2--1 and $^{13}$CO $J$~=~1--0 (hereafter $R^{13}_{\rm 2-1/1-0}$) {in order to derive} the physical condition of the {molecular clouds.} 
The CO line intensity ratio between different rotational excitation levels reflects the kinetic temperature and density of the molecular gas.
The spatial resolution of the $^{13}$CO $J$~=~2--1 and $^{13}$CO $J$~=~1--0 data are smoothed to be \timeform{120"}\ to improve signal-to-noise {ratios}.
The grid size of $^{13}$CO $J$~=~2--1 data obtained by APEX is aligned to $^{13}$CO $J$~=~1--0 data taken from Mopra by using the miriad software.
Figure \ref{ratio} shows the $R^{13}_{\rm 2-1/1-0}$ map. The intensity ratio map was created above $1.7$ K \kms in the $^{13}$CO $J$~=~2--1 and $^{13}$CO $J$~=~1--0 integrated intensities.
$R^{13}_{\rm 2-1/1-0}$ is enhanced in the head and tail of Cloud 1 to be $R^{13}_{\rm 2-1/1-0}\sim 0.9$ and $R^{13}_{\rm 2-1/1-0}\sim 0.85$, while it has low of $R^{13}_{\rm 2-1/1-0}\sim 0.5$ around $(l,b)=(\timeform{331.67D}$, $\timeform{0.02D}$).
In the head of Cloud 2, the ratio is enhanced to be $R^{13}_{\rm 2-1/1-0}\sim 1.1$.
\begin{figure*}[h]
\begin{center} 
 \includegraphics[width=18cm]{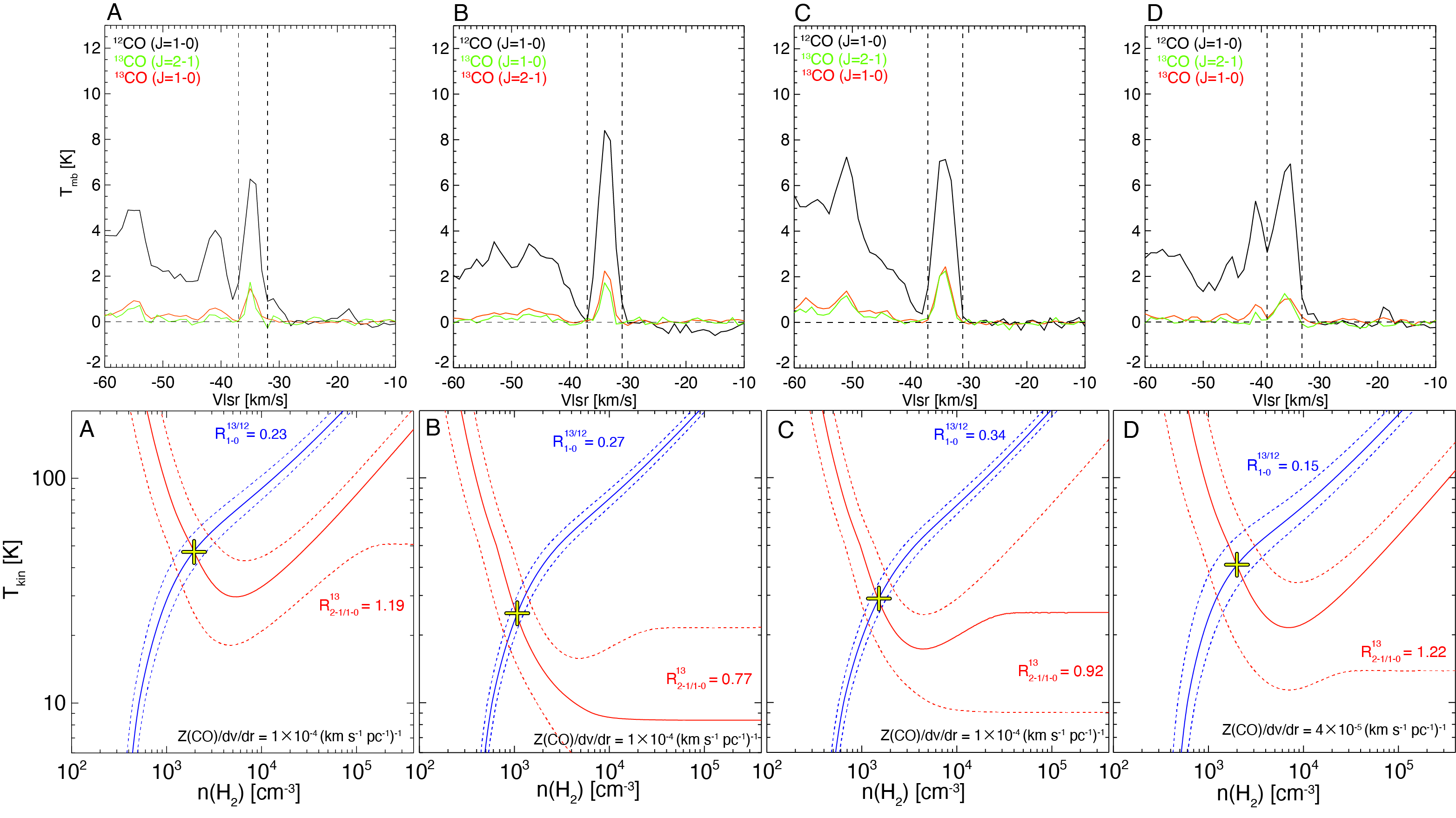}
\end{center}
\caption{Spectra and results of the LVG calculation at the point A, B, C, and D. The black, green, and red spectra show the $^{12}$CO $J$~=~1--0, $^{13}$CO $J$~=~2--1, and $^{13}$CO $J$~=~1--0, respectively. The vertical dotted lines indicate the integrated velocity range. Blue and red curves of constant $R^{13/12}_{\rm 1-0}$ and $R^{13}_{\rm 2-1/1-0}$ as functions of the molecular gas density [$n({\rm H_2})$] and kinetic temperature [$T_{\rm kin}$]. {$Z({\rm CO})$ is the CO and H$_2$ abundance ratio given by $[^{12}{\rm CO}]/[{\rm H_2}]=10^{-4}$ \citep{1982ApJ...262..590F,2010ApJ...721..686P}}. The dotted lines show the $\pm 1\ \sigma$ error of the intensity ratio. The yellow cross marks indicate the solution of each data point. Alt text: Spectra and results of LVG calculation at the point A, B, C, and D.}
\label{lvg}
\end{figure*}

\begin{table*}
\caption{Results of the LVG analysis of the head-tail molecular clouds.} 
\begin{center}
\begin{tabular}{ccccccccc} 
\hline
Name  & $l$ & $b$ &$\Delta V$ & $r$ &$R^{13}_{\rm 2-1/1-0}$ &$R^{13/12}_{\rm 1-0}$ & $n({\rm H_2})$ &$T_{\rm kin}$\\
  & [deg] & [deg] & [\kms] & [pc] & & & [cm$^{-3}$] &[K]\\
 (1) & (2) & (3) & (4) & (5) & (6) & (7)& (8)& (9)\\
\hline
A & 331.719 & $+0.240$& 2.2 & 2.7 &1.19&0.23&$\left(2.0^{+1.0}_{-0.7}\right)\times 10^3$&  $47^{+21}_{-18}$  \\
B & 331.660 &$+0.073$ &2.1& 2.1 &0.77&0.27&$\left(1.1^{+0.4}_{-0.3} \right)\times 10^3$& $25^{+13}_{-9}$  \\
C & 331.619 & $-0.069$ & 2.8 & 1.8 &0.92&0.34&$\left(1.5^{+0.6}_{-0.4}\right)\times 10^3$&  $29^{+13}_{-10}$ \\
D & 331.844 & $+0.190$& 3.8 &1.7&1.22&0.15&$\left(2.0^{+1.3}_{-0.9}\right)\times 10^3$& $41^{+26}_{-19}$ \\
\hline
\end{tabular} 
\label{lvgtable}\\
\end{center}  
{\raggedright Note. (1) Position name (2) Galactic longitude (3) Galactic latitude (4) FWHM of the $^{13}$CO $J$~=~1--0 spectra. (5) Radius defined by $\sqrt{S/\pi}$ above the 30\% level of the peak $^{13}$CO $J$~=~1--0 integrated intensity. The boundary between B and C is decided at $b=0$. (6) Intensity ratio between $^{13}$CO $J$~=~2--1 and $^{13}$CO $J$~=~1--0. (7) Intensity ratio between $^{13}$CO~$J$~=~1--0 and $^{12}$CO~$J$~=~1--0. (8) The number densities of molecular hydrogen. (9) The kinetic temperatures. 
\par}
\end{table*} 

We calculated the kinetic temperature [$T_{\rm kin}$] and {molecular} hydrogen number density [$n({\rm H_2})$] {by} assuming the Large Velocity Gradient (LVG) model \citep{1974ApJ...189..441G,1974ApJ...187L..67S}. The LVG calculation assumes a spherical molecular cloud with a constant velocity gradient ($dv/dr$), {and uniform density} and temperature. It solves the statistical equilibrium equations at each rotational energy level by considering the effect of optical depth with the photon escape probability \citep{1970MNRAS.149..111C}.

Figure \ref{lvg} shows the spectra and results of the LVG {calculations} for data points of A, B, C, and D as shown in Figure \ref{ratio}. The calculations used the integrated intensity ratio in the velocity range shown by the vertical dotted lines of the spectrum. We assumed that the CO and H$_2$ abundance ratio is {given as} $[^{12}{\rm CO}]/[{\rm H_2}]=10^{-4}$ \citep{1982ApJ...262..590F,2010ApJ...721..686P} and the isotope abundance ratio is [$^{12}$C]/[$^{13}$C] $= 53$ (4 kpc ring in Table 4 of \cite{1994ARA&A..32..191W}). 
The velocity gradient $(dv/dr)$ was calculated from the FWHM and {cloud} radius ($r$). FWHM is estimated {by a} single Gaussian fitting to each $^{13}$CO $J$~=~1--0 {spectrum}. $r$ is derived by $\sqrt{S/\pi}$, where $S$ is the cloud area above the 30\% level of the peak integrated intensity. {By the analysis we derived that} Cloud 1 and Cloud 2 have $T_{\rm kin} = 25$--$50$ K {and} $n({\rm H_2}) = (1.0$--$2.0) \times 10^3$ cm$^{-3}$.
In particular, the head and tail parts (A, C, and D) {have high temperatures as compared with} 10 K, the typical kinetic temperature of molecular gas without {extra} heating sources (e.g., \cite{2015ApJS..216...18N}).
The results of the LVG calculations at each point A, B, C, and D are summarized in Table \ref{lvgtable}.


\subsection{Comparison with the infrared image}
We compared the CO data with Herschel infrared images to investigate the origin of the heating in the head-tail molecular clouds. 
Figure \ref{infra}(a) shows the spatial distribution of the head-tail molecular clouds overlaid on the Herschel 160 $\mu$m continuum image \citep{2010PASP..122..314M}. 
Figure \ref{infra}(b) presents a close-up map image of the head part in Cloud 1. 
The infrared peak close to the tip of Cloud 1 is the massive star-forming region G331.5-0.1 associated with the Norma arm at a distance of 7.5 kpc (\cite{2013ApJ...774...38M}), {which exists at the background source toward the head-tail molecular clouds as discussed in Section 4.1.} 
{So, we find no infrared source toward Cloud 1 and  Cloud 2 which have high kinetic temperatures,} $T_{\rm kin} = 47^{+21}_{-18}$ K of and $T_{\rm kin} = 41^{+26}_{-19}$ K, respectively. {This indicates that there is no responsible radiative heating source {for the high temperature} in the region of Cloud 1 and Cloud 2.} 


\begin{figure*}[h]
\begin{center} 
 \includegraphics[width=16cm]{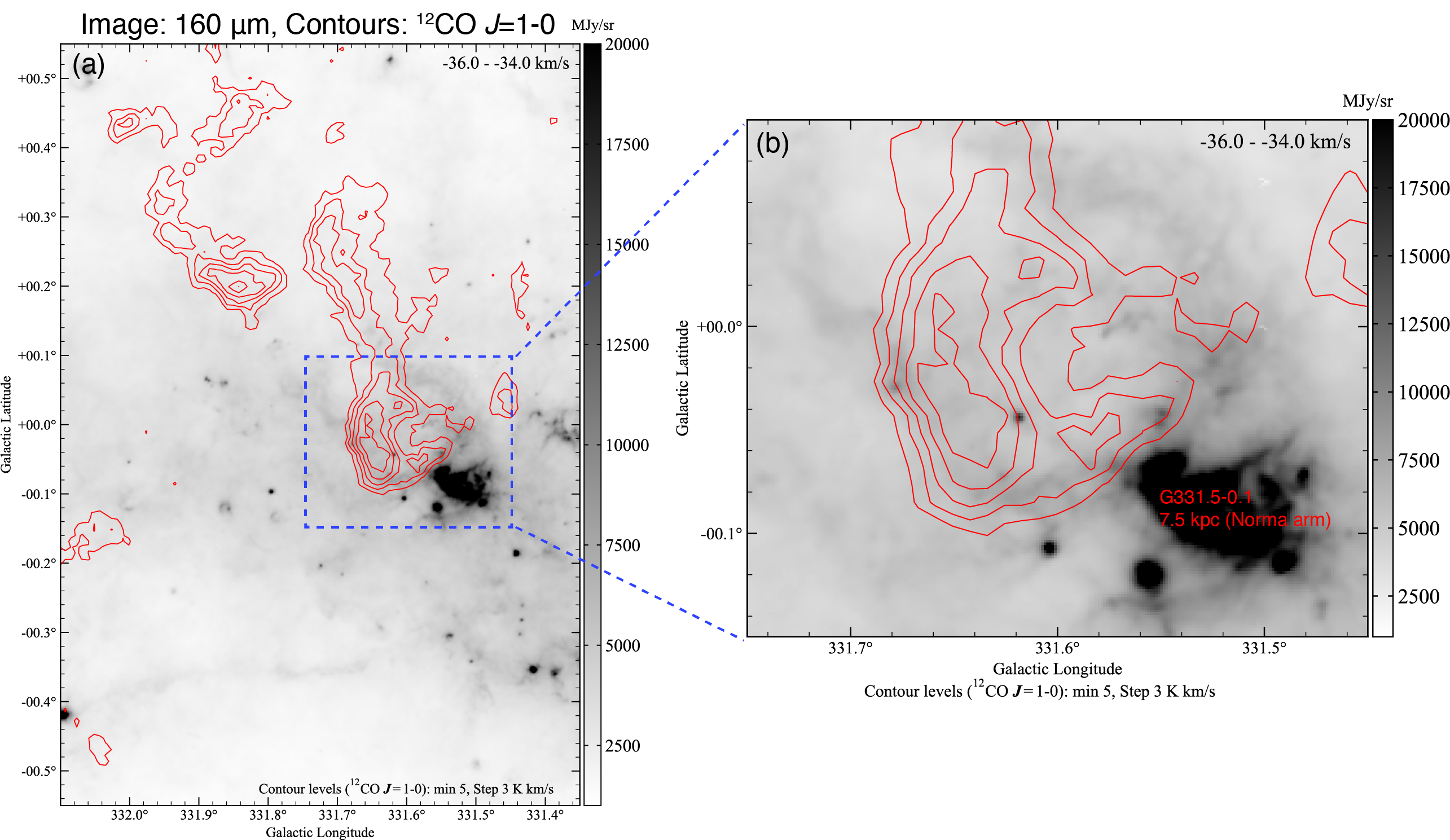}
\end{center}
\caption{(a) $^{12}$CO $J$~=~1--0 spatial distributions of head-tail molecular clouds superposed on the Herschel 160 $\mu$m continuum image \citep{2010PASP..122..314M}. The lowest contour levels and intervals are 5.0 K \kms and 3.0 K \kms. (b) The close-up image of panel (a). Alt text: Mopra 12-CO spatial distributions of head-tail molecular clouds superposed on the Herschel 160 mircrometer continuum image.}
\label{infra}
\end{figure*}



\section{Discussion}
\subsection{Distance to the head-tail molecular clouds}
\begin{figure*}[h]
\begin{center} 
 \includegraphics[width=18cm]{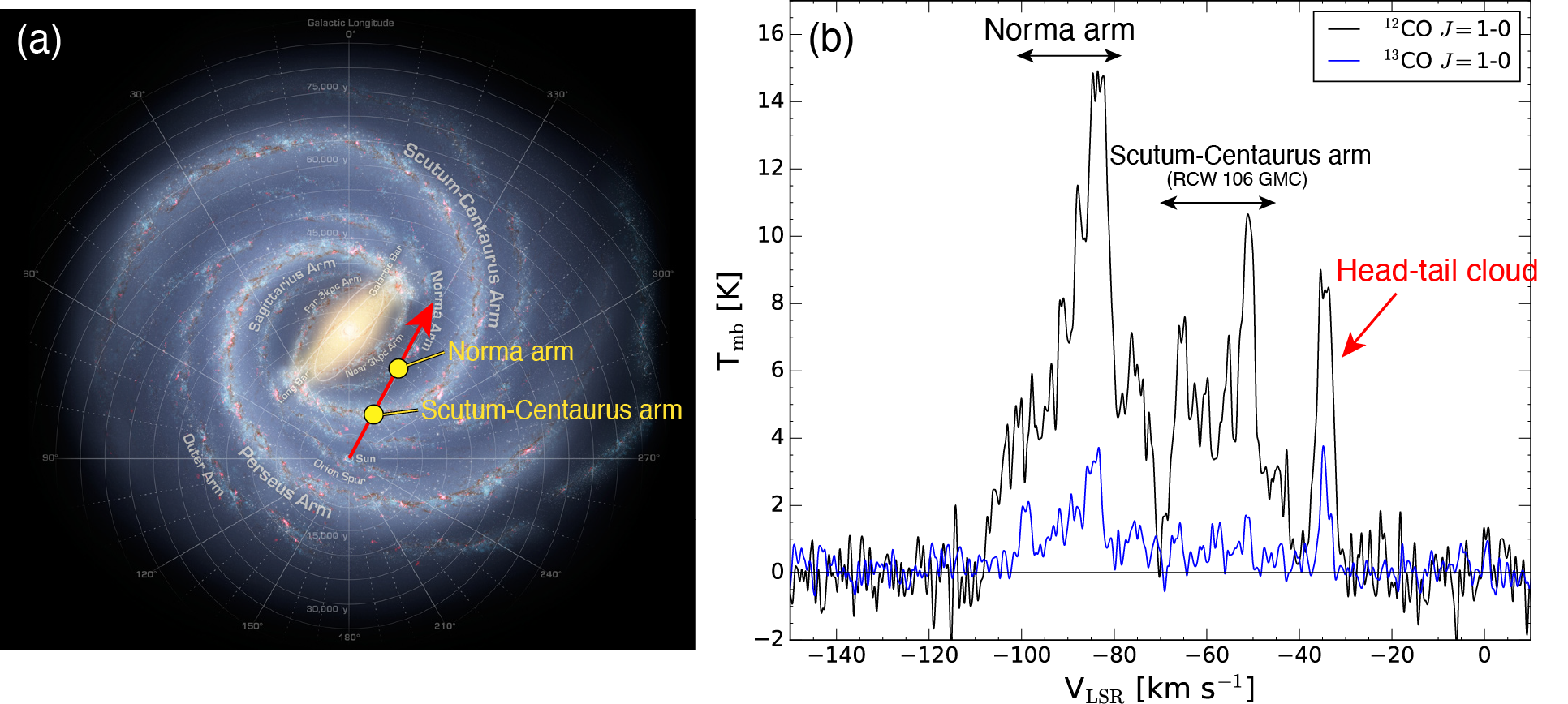}
\end{center}
\caption{(a) The top schematic view of the Milky Way (NASA/JPL-Caltech/ESO/R. Hurt). The red arrow indicates the direction of $l=\timeform{331D}$. Two yellow circles are present at the intersections of the line-of-sight with the Scutum-Centaurus and Norma arms. (b) Spectra of Mopra $^{12}$CO and $^{13}$CO $J$~=~1--0 at $(l,b)=(\timeform{331.627D}, \timeform{-0.0604D})$. Alt text: The top schematic view of the Milky Way and spectra of Mopra 12-CO and 13-CO.}
\label{spec}
\end{figure*}
{In order to investigate the properties of the present clouds, we need a reliable distance estimate.}
Figure \ref{spec}(a) shows a face-on schematic view of the Milky Way, and the red arrow indicates the line of sight at $l= \timeform{331D}$.
The Norma arm and Scutum-Centaurus arm overlap in this direction.
Figure \ref{spec}(b) shows the spectrum at the head of Cloud 1.
The velocity component from $-100$ \kms to $-80$ \kms corresponds to the Norma arm, which is associated with the massive star-forming region G331.5-0.1 (\cite{2013ApJ...774...38M}). 
The velocity component from $-80$ \kms to $-50$ \kms is associated with the RCW 106 (G333) GMC complex at the Scutum-Centaurus arm \citep{2015ApJ...812....7N,2023A&A...676A..69Z,2025AJ....169..181K}.
The head-tail molecular clouds have $-35$ \kms, {which is red-shifted by more than 15 \kms from these spiral arms and {do} not correspond to the main velocity component in the spiral arms in the direction of $l=\timeform{331D}$ \citep{2017AstRv..13..113V}. }

\begin{figure*}[h]
\begin{center} 
  \includegraphics[width=12cm]{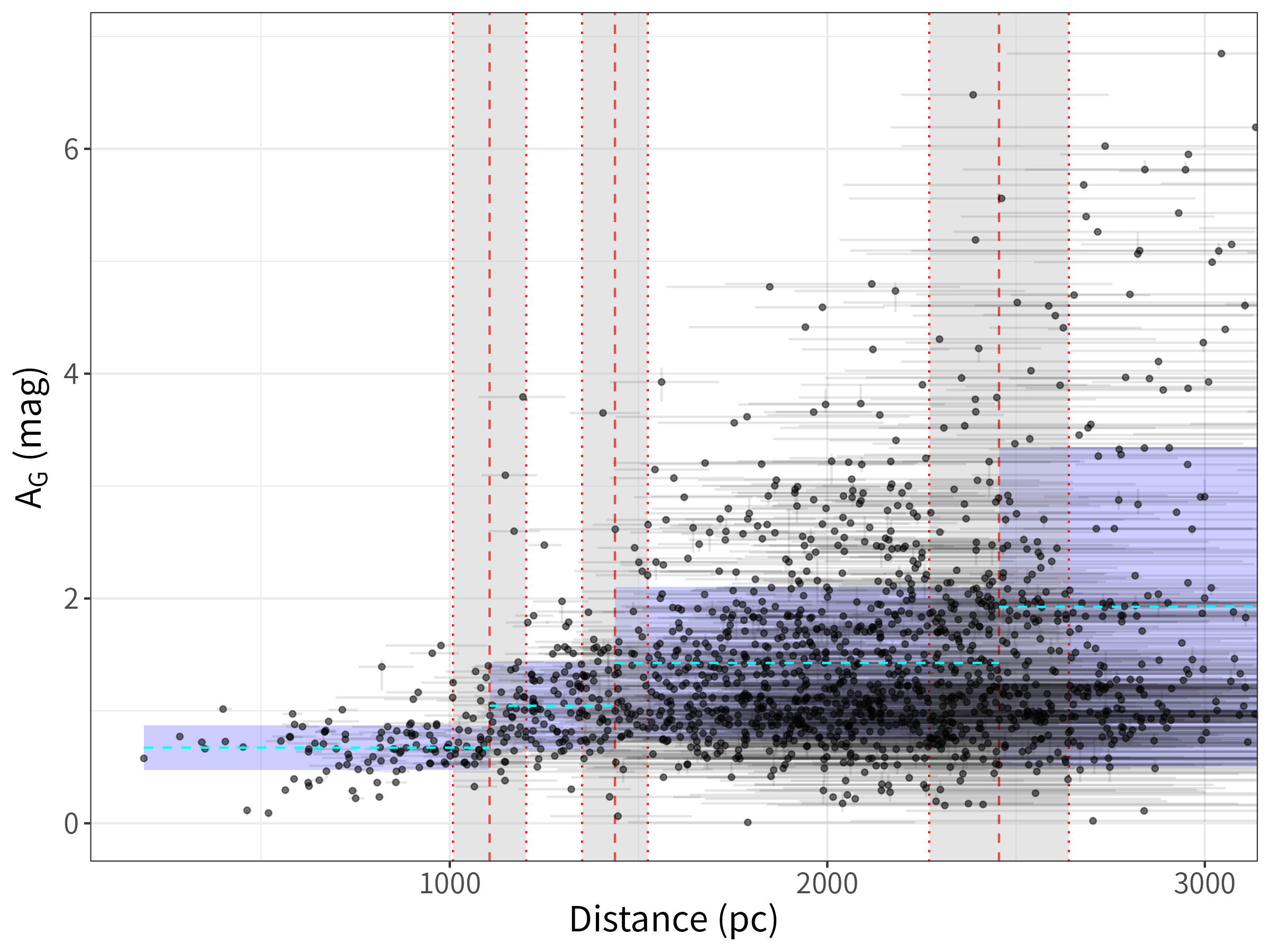}
\end{center}
\caption{The scatter plot between $A_{\rm G}$ and the distance of fixed stars corresponding to Cloud 1. The vertical red bold lines indicate the breakpoints of $A_{\rm G}$. The vertical grey shadow areas show the ranges of $\pm 1\sigma$ standard deviation in each break point. The horizontal light blue dotted line and purple shadow area show the median and mean absolute deviation ranges of $A_{\rm G}$ divided into three break points, respectively. The mean absolute deviation values are scaled to $1\sigma$ standard deviation. The uncertainties of $A_{\rm G}$ are their 16th and 84th percentiles from the Gaia DR3 catalog \citep{2023A&A...674A...1G}. The errors of the geometric distance are their 16th and 84th percentiles, which correspond to a 68\% confidence interval \citep{2021AJ....161..147B}. Alt text: The scatter plot between G-band extinction and the distance of fixed stars corresponding to Cloud 1 obtained by Gaia DR3.}
\label{gaiadistance}
\end{figure*}

In order to estimate the distance to the head-tail clouds, we analyzed the data of fixed stars obtained by Gaia Data Release 3 \citep{2023A&A...674A...1G}.
In this paper, we used a stellar distance catalog based on \citet{2021AJ....161..147B} to attempt to estimate the distance of the head-tail molecular clouds from the G-band extinction ($A_{\rm G}$) and annual parallax measurements.
Figure \ref{gaiadistance} presents scatter plots between $A_{\rm G}$ and the distance of Gaia sources corresponding to Cloud 1, {which has the $^{12}$CO $J$~=~1--0 integrated intensity greater than 5 K \kms.}
If {the} interstellar dust is distributed uniformly, $A_{\rm G}$ increases with distance. {However} there is a dense molecular cloud, $A_{\rm G}$ has a breaking point. 
Thus, we can estimate the line-of-sight distance of the molecular cloud, {by} analyzing the breaking points of $A_{\rm G}$ (e.g., \cite{2019A&A...624A...6Y,2019ApJ...879..125Z,2021ApJ...914..122D,2024ApJ...961...13D}).
Break point analysis methods that apply to Gaia DR3 in this paper were presented by \citet{2021ApJ...914..122D} and in Appendix C of \citet{2024ApJ...961...13D}.
Based on the analysis of the break points, we found that there are three break points at distances of $1.11\pm0.09$ kpc, $1.44 \pm 0.09$ kpc, and $2.46 \pm 0.18$ kpc, as shown by the vertical red dotted lines in Figure \ref{gaiadistance}. In order to determine the distance, we investigated the spatial distribution of Gaia sources at these three breaking points.

\begin{figure*}[h]
\begin{center} 
  \includegraphics[width=17cm]{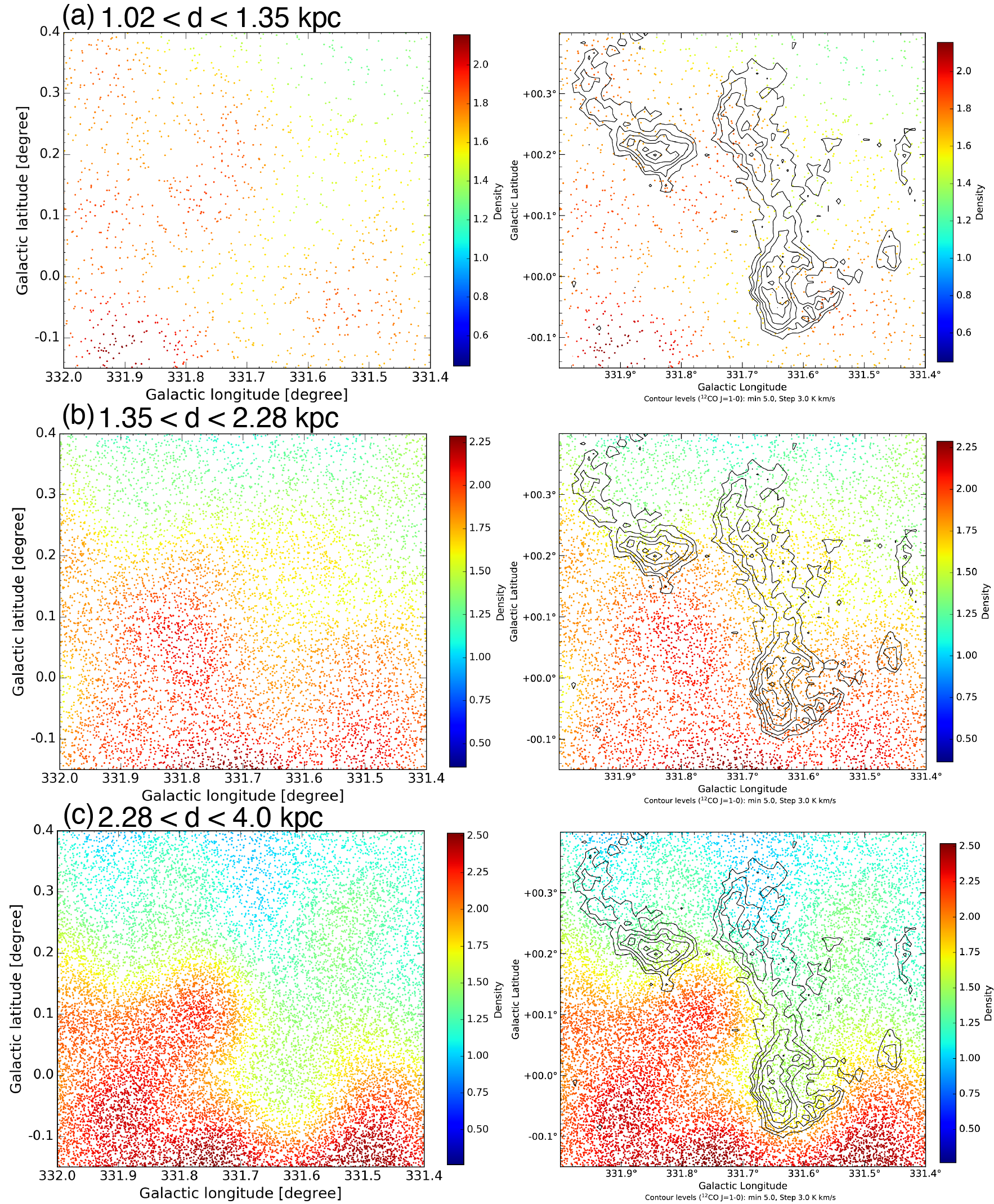}
\end{center}
\caption{The stellar density map of fixed stars obtained by Gaia DR3. The panels (a), (b), and (c) show from 1.02 kpc to 1.35 kpc, from 1.35 kpc pc to 2.28 kpc, and from 2.28 kpc pc to 4.0 kpc, respectively. The black contours in the right column are the same as Figure \ref{mopra} (a). Alt text: The stellar density map of fixed stars obtained by Gaia DR3.}
\label{gaia}
\end{figure*}

Figures \ref{gaia} (a), (b), and (c) show stellar density maps of Gaia sources at distances from 1.02 kpc to 1.35 kpc, from 1.35 kpc to 2.28 kpc, and from 2.28 kpc to 4.0 kpc, respectively. 
The right panels of each figure show $^{12}$CO~$J$~=~1--0 contours overlaid on the stellar density map. 
The boundaries at 1.02 kpc, 1.35 kpc, and 2.28 kpc correspond to the distance of the lower limit at the three breaking points.
We cannot find density distributions corresponding to {Cloud 1} in the panels at distances from 1.02 kpc to 1.35 kpc and from 1.35 kpc to 2.28 kpc as shown in Figure \ref{gaia} (a) and (b).
On the other hand, in the panel from 2.28 kpc to 4.0 kpc, the extinction of stellar density has morphological correspondence with CO distributions at the tips of {Cloud 1} (Figure \ref{gaia}c).
These results indicate that {Cloud 1} exists at a distance within a range from 2.28 kpc to 4 kpc.
Thus, we suggest that $2.46\pm 0.18$ kpc is the most likely distance to {Cloud 1} among the three breaking points. 
If we adopt distance $d = 2.46$ kpc, the length of {Cloud 1 in $^{12}$CO $J$~=~1--0 is $\sim 15$ pc and its width is $\sim 5$ pc.} The total molecular mass ($M_{\rm cloud}$) is given by
\begin{equation}
M_{\rm cloud} = \mu_{\rm H_2} m_{\rm H} d^2 \Omega \sum_{i} N_i(\rm{H_2})\ [M_{\odot}],
\label{mass}
\end{equation}
where $\Omega$ is the solid angle subtended by the molecular cloud, $\mu_{\rm H_2}$ the average molecular weight of a hydrogen molecule, 2.8, $m_{\rm H}$ the proton mass, $1.67 \times 10^{-24}$ g, and $N_i(\rm{H_2})$ the hydrogen molecular column density at the $i$-th pixel. The molecular cloud {mass of Cloud 1 is} calculated as $4.8\times 10^3\ M_{\odot}$.
{The CO distributions of Cloud 2 also seem to fit the stellar density map from 1.35 kpc to 2.28 kpc as well as from 2.28 kpc pc to 4.0 kpc. On the other hand, the radial velocities of Cloud 1 and Cloud 2 are in a common velocity range from $-36$ \kms to $-34$ \kms.
This result shows that Cloud 2 is likely to be {at} the same distance $2.46 \pm 0.18$ kpc estimated {for} Cloud 1, while we cannot exclude that Cloud 2 possibly exists with a foreground distance of $1.44 \pm 0.09$ kpc. 
If we assume 2.46 kpc to the distance of Cloud 2, its length, width, and total mass are derived to be $\sim 10$pc, $\sim 3$ pc, and $3.5 \times 10^3\ M_{\odot}$. We note that the difference in the possible distance of Cloud 2 does not affect the falling cloud scenario onto the Milky Way disk that we discuss in the following sections.}
\subsection{Interaction of the head-tail molecular clouds with the gas disk component}
\begin{figure*}[h]
 \includegraphics[width=18cm]{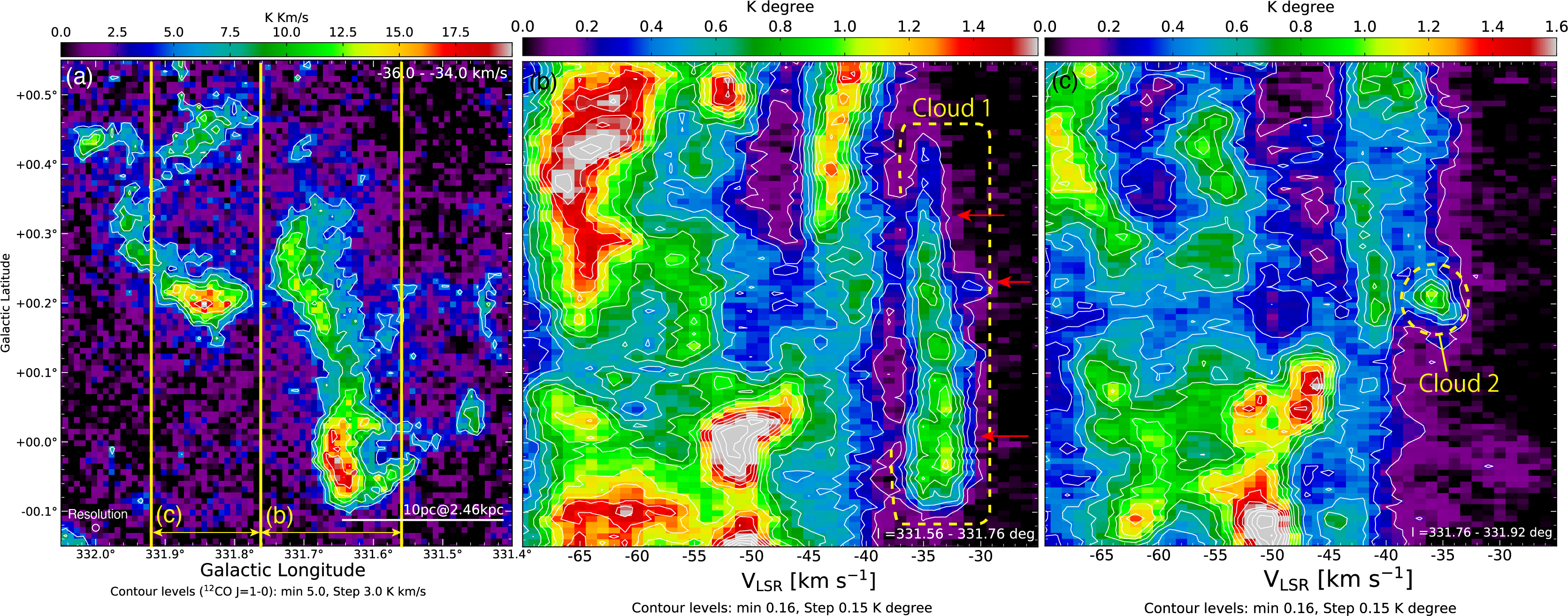}
\caption{(a) Same as Figure \ref{mopra}(a). The yellow lines show the integrated longitude ranges in the panels (b) and (c). (b) The $^{12}$CO $J$~=~1--0 latitude-velocity diagram of Cloud 1. The red arrows indicate the bridge features connecting to the disk velocity component. The integrated longitude range is from \timeform{331.56D} to \timeform{331.76D}. (c) The $^{12}$CO $J$~=~1--0 latitude-velocity diagram of Cloud 2. The integrated longitude range is from \timeform{331.76D} to \timeform{331.92D}. The lowest contour levels and intervals are 0.16 K degree and 0.15 K degree, respectively. Alt text: Mopra 12CO integrated intensity map, latitude-velocity diagram of Cloud 1, and latitude-velocity diagram of Cloud 2.}
\label{bv}
\end{figure*}

We present position-velocity diagrams toward Cloud 1 and Cloud 2 in order to investigate the relationship with the gas disk components. 
Figure \ref{bv}(a) shows the spatial distribution of the head-tail molecular clouds. {Figures \ref{bv}(b) and (c) present} latitude-velocity diagrams of Cloud 1 and Cloud 2, respectively. \citet{2025AJ....169..181K} reported the three velocity components at $-64$ \kms, $-54$ \kms, and $-44$ \kms toward RCW 106 West, which includes the region analyzed in {the present} paper (see Figures 9 and 10 in \cite{2025AJ....169..181K}). 
The authors suggested that the $-44$ \kms component with a velocity range from $-50$ \kms to $-38$ \kms is a diffuse foreground component different from the RCW 106 GMC complex at a distance of 3.6 kpc, and exists at the front face of the Scutum--Centaurus arm. A previous study of \citet{2005A&A...429..497R} also supports {the interpretation of the diffuse component} based on a velocity-resolved deep H$\alpha$ survey. 
Cloud 1 component connects to the $-44$ \kms component with the bridge features as shown by red arrows in Figure \ref{bv}(b).
We also find Cloud 2 as a compact peak in Figure \ref{bv}(c), which merges with the $-44$ \kms component. 
The numerical simulations of cloud-cloud collisions (\cite{2014ApJ...792...63T}, \yearcite{2018PASJ...70S..58T}) reproduce bridge features and merging two velocity components in a position-velocity diagram of the {collision} between two molecular clouds (see also Figure 9 in \cite{2015MNRAS.454.1634H} and Figure 10 in \cite{2017ApJ...835..142T}).   
Based on these results, we propose that head-tail molecular clouds interact with the Galactic gas disk component at the front face of the Scutum--Centaurus arm. 

\subsection{The falling scenario onto the Milky Way disk}
{We first discuss the morphological properties of the clouds.}
We suggest that the morphology of the head-tail molecular clouds is explained by falling motion onto the Galactic disk.
Indeed, the magnetohydrodynamic (MHD) simulations were reproduced with the head-tail morphology toward the Galactic plane by falling motion onto the disk from the Galactic halo (see Figure 7 in \cite{2009ApJ...699.1775K},{\cite{2022ApJ...925..190S}}). 
{The former simulation shows that the falling cloud has a U-shaped shocked region with enhanced} density and turbulence in the tip of the head.
{These are similar properties to the present clouds, and we propose that the origin of the heating and large velocity dispersion at the head of head-tail molecular clouds were caused by the shock compression and heating by the interactions with the Galactic disk. }
We point out that the head-tail cloud in their simulation by \citet{2009ApJ...699.1775K} having a size of $\sim 3$ kpc from the head to the tail. This size is a factor $\sim 200$ larger than the head-tail molecular cloud reported in this paper, but their results are useful to infer the correspondence of morphologies between observation and simulation since the local shock interactions are not dominated by gravity {and are scale-free}. More realistic simulations of smaller molecular clouds in the future will be useful in understanding more detailed correspondence in temperature, etc.

{The origin of the falling clouds needs to be explored. As discussed in the introduction, in the 1960’s HI observations revealed the high-velocity clouds (HVCs) and the intermediate velocity clouds (IVCs) at high Galactic latitudes have radial velocity of $|V_{\rm LSR}| \gtrsim 50$-$100$ \kms without rotation }(\cite{1963CRAS..257.1661M,1963BAN....17..203S,1970A&A.....7..381O,1997ARA&A..35..217W,1998MNRAS.299..611B,1999ApJ...514..818B,2001ApJS..136..463W,2004PASJ...56..633S,2015MNRAS.447L..70F,2008ApJ...679L..21L};{\cite{2025SciA...11S4057L}}). 
{It is possible that the present falling clouds are linked with the IVCs because their velocities are in a similar range.} The origins of the IVCs are thought to be the accretion of gas ejected from the Galactic disk into the halo by the feedback of supernova explosions called Galactic Fountains  (\cite{1980ApJ...236..577B,2013MNRAS.433.1634M}), or low-metallicity atomic gas falling into the Galactic disk from outside the Galaxy (\cite{2021PASJ...73S.117F,2022arXiv220813406H}). {In particular, it is worth noting that the absorption line measurements supporting the Galactic Fountains were made in only several lines of sight toward the bright background stars/galaxies in the IVCs. This makes a strong contrast with the observations of dust-to-gas ratio made by \citet{2022arXiv220813406H}, which covers consecutively the whole IVCs at $b$ higher than 15 degrees. We argue that the low metallicity IVCs from the outside the Milky Way are falling on to the Milky Way disk and might be forming to molecular gas like head-tail molecular clouds presented in this paper. }

{The recent study of an IVC named IVC 86-36 in PP Arch has a head-tail morphology in the radial velocity from $-50$ to $-30$ \kms, and the total cloud mass of $\sim 7 \times 10^3\ M_{\odot}$ \citep{2021PASJ...73S.117F}. The mass is comparable to those of Cloud 1 and Cloud 2, while the size of the head of IVC 86-36 is $\sim 90$ pc, an order of magnitude larger than the present clouds.}
{\citet{2022ApJ...925..190S} showed that the observational features of IVC 86-36 can be explained by collision and merging with the Galactic halo-disk from their numerical simulations. We therefore propose that the head-tail molecular clouds are falling onto the Galactic disk in a similar way to the IVC 86-36, while a process which converts the HI gas into molecular form is required to explain the CO clouds.}
{It is possible that the conversion {of HI into H$_2$} occurs at a lower height {of} $\ll 1$ kpc from the plane, where the ambient pressure due to {the} HI gas disk can become high enough to compress HI to high density more than $10^3$ cm$^{-3}$. At the density, HI can be converted to H$_2$ via dust surface reaction in Myr (e.g., \cite{1971ApJ...163..155H}). It is also to be noted that the estimated temperature of the clouds is as high as 30-50 K. This suggests that the shock heating is fairly strong, involving high-speed collision at a relative velocity higher than 35 \kms, which is consistent with the IVC velocity. }


A few other examples of a head-tail cloud with a head-tail distribution were discovered in the N159 region of the Large Magellanic Cloud (LMC), where the clouds likely originated from high-velocity HI gas colliding onto the LMC disk at $>50$ \kms (\cite{2019ApJ...886...14F,2019ApJ...886...15T}, \yearcite{2022ApJ...933...20T}). These clouds were interpreted based on the MHD simulations of a falling cloud by \citet{2018PASJ...70S..53I} (see also for visualization of the results \cite{2024arXiv240806826M}). 
{The clouds show heated head regions, while the heating was made radiatively by the high mass stars {formed instead of the shock heating.} The observed shock velocity {in N159} is $\sim 20$ \kms, {probably} somewhat lower than the case of the IVC origin, and the shock heating does not dominate the heating. }
{The discovery of the falling CO clouds has revealed for the first time that dense CO clouds are falling down to the Galaxy at high velocity.
{While detections of CO lines in IVCs were reported previously  (e.g., \cite{1990ApJ...355L..51D,1999A&A...344..955W,2010ApJ...722.1685M,2016A&A...592A.142R}), direct observational signatures of falling motion such as the moving direction and velocity were not revealed. The discovery offers a tight connection between the Galactic halo and disk, and can connect the Milky Way and the inter-galactic matter.}
The discovery raised the following issues to be addressed in future studies;
\begin{itemize}
\item  Are the falling CO clouds a common phenomenon in the Galaxy? If yes, what is the fraction of the total mass included in the falling CO clouds.
\item {Whether the origin of the falling clouds is the HI gas in the IVCs, and whether the falling motion follows that of the IVCs.}
\item Are the falling CO clouds forming stars? What are the impacts of the falling clouds on star formation? 
\item What is the role of the falling clouds in cycling the ISM in the Galaxy?
\end{itemize}
Observational and theoretical studies on these issues will be beneficial to better understand the galaxy evolution in a three-dimensional perspective, while the conventional perspective is mainly limited to a thin layer of the Galactic plane. The three-dimensional {view} may suggest the Galaxy as a system more open to the intergalactic space.}

\clearpage
\section{Conclusion}
The conclusions of this paper are as follows:
\begin{enumerate}
\item Two head-tail molecular clouds were identified at $l=\timeform{331.6D}$ {and $b=\timeform{0.0D}$.}
A detailed analysis of the CO data obtained by Mopra revealed that the {two head-tail molecular clouds have the radial velocity of $-35$ \kms}, which is different from the velocity components of the Scutum--Centaurus and Norma arms of the Milky Way in the line of sight at $l=\timeform{331.6D}$.
\item The kinetic temperatures of the head-tail clouds were estimated from an analysis of the emission line intensity ratio of the APEX $^{13}$CO~$J=$~2--1 and the Mopra $^{13}$CO $J$~=~1--0 data. The heads and tail parts were $T_{\rm kin}=30$-$50$ K, which is more heated than the quiescent molecular gas without heating sources. A comparison with the Herschel 160 $\mu$m data revealed no infrared source as heating sources corresponding to the entire head-tail molecular clouds.
\item We find a G-band extinction corresponding to the head-tail molecular cloud by analyzing the Gaia Data Release 3. Based on the breaking points analysis and morphological correspondence of interstellar extinction, the head-tail molecular clouds are likely to exist at a distance of $2.46\pm 0.18$ kpc in the foreground of the RCW 106 GMC complex. The masses of Cloud 1 and Cloud 2 are derived to be $4.8\times 10^3\ M_{\odot}$ and $3.5\times 10^3\ M_{\odot}$, respectively.
\item We suggest that these head-tail moleclouds are falling clouds onto the Milky Way disk, and are heated by the shock compression with the interaction of the gas disk component. {The shock velocity inferred is more than $35$ \kms for the typical observed velocity of the IVCs, 40-60 \kms, which explains the high temperatures of the shocked gas in the present falling clouds.} {The present work revealed the falling molecular clouds for the first time based on direct observational signatures, which include the moving direction toward the plane by the head-tail morphology and high velocity similar to the IVCs. }
\end{enumerate}

\section*{Acknowledgements}
{We are grateful to the anonymous referee for carefully reading our manuscript and giving us thoughtful suggestions, which greatly improved this paper.}

We utilized the Astropy \citep{2013A&A...558A..33A,2018AJ....156..123A,2022ApJ...935..167A}, NumPy \citep{2011CSE....13b..22V}, Matplotlib \citep{2007CSE.....9...90H}, IPython \citep{2007CSE.....9c..21P}, Miriad \citep{1995ASPC...77..433S}, and APLpy \citep{2012ascl.soft08017R}.

The authors are grateful to Dr. Michael Burton of the University of New South Wales and the Armagh Observatory and Planetarium for the archival CO survey data with Mopra.
We also thank Dr. Graeme Wong for kindly supporting remote observations from Nagoya University.

{This work was supported by Grants-in-Aid for Scientific Research (KAKENHI) from Japan Society for the Promotion of Science (JSPS), Grant Numbers JP15H05694, JP20H01945, JP21H00040, JP22H00152, and JP24H00246.}
The Mopra radio telescope is part of the Australia Telescope National Facility, which is funded by the Australian Government for operation as a National Facility managed by CSIRO. The University of New South Wales Digital Filter Bank used for the observations with the Mopra Telescope was provided with support from the Australian Research Council.

This publication is based on data acquired with the Atacama Pathfinder Experiment (APEX) under programmes 092.F-9315 and 193.C-0584. APEX is a collaboration among the Max-Planck-Institut fur Radioastronomie, the European Southern Observatory, and the Onsala Space Observatory. The processed data products are available from the SEDIGISM survey database located at \url{https://sedigism.mpifr-bonn.mpg.de/index.html}, which was constructed by James Urquhart and hosted by the Max Planck Institute for Radio Astronomy

The Herschel spacecraft was designed, built, tested, and launched under a contract to ESA managed by the Herschel/Planck Project team by an industrial consortium under the overall responsibility of the prime contractor Thales Alenia Space (Cannes), and including Astrium (Friedrichshafen) responsible for the payload module and for system testing at spacecraft level, Thales Alenia Space (Turin) responsible for the service module, and Astrium (Toulouse) responsible for the telescope, with in excess of a hundred subcontractors.

This work presents results from the European Space Agency (ESA) space mission Gaia. Gaia data are being processed by the Gaia Data Processing and Analysis Consortium (DPAC). Funding for the DPAC is provided by national institutions, in particular the institutions participating in the Gaia MultiLateral Agreement (MLA). The Gaia mission website is \url{https://www.cosmos.esa.int/gaia}. The Gaia archive website is \url{https://archives.esac.esa.int/gaia}.





\end{document}